\documentclass[fleqn,12pt,twoside]{article}
\usepackage{espcrc1}

\usepackage{graphicx}
\usepackage[figuresright]{rotating}

\newcommand{\AmS}{{\protect\the\textfont2
  A\kern-.1667em\lower.5ex\hbox{M}\kern-.125emS}}

\title{Pygmy Dipol Resonances as a Manifestation of the Structure of the Neutron-Rich Nuclei}

\author{N. Tsoneva\address[GI]{Institut f\"ur Theoretische Physik, Universit\"at
Gie\ss en, Heinrich-Buff-Ring 16, \\
D-35392 Gie\ss en, Germany}
        \thanks{Also at Institute for Nuclear Research and Nuclear Energy, 1784 Sofia, Bulgaria},
        H. Lenske\addressmark[GI]
        and
        Ch. Stoyanov\address{Institute for Nuclear Research and Nuclear Energy, 1784 Sofia, Bulgaria}}

\begin{document}       
\maketitle

\begin{abstract}
Dipole excitations in neutron-rich  nuclei below the neutron threshold are investigated. 
The method is based on Hartree-Fock-Bogoliubov (HFB) and Quasiparticle-Phonon Model (QPM) theory. Of our special interest are the properties of the low-lying 1$^-$ {\it Pygmy Resonance} and the two-phonon quadrupole-octupole $1^{-}$
states in Sn-isotopes including exploratory investigations for the experimentally unknown mass regions. In particular we investigate the evolution of the dipole strength function with the neutron excess. The use of  HFB mean-field potentials and s.p. energies is found to provide a reliable extrapolation into the region off stability. 
\end{abstract}

\section{\label{}Introduction \protect }

The understanding
of the structure of atomic nuclei at extreme isospin is a challenge for nuclear
physics which has become achievable on a large scale after the modern experimental nuclear structure facilities \cite{Han} came into operation.
For the theoretical studies of such systems of importance is the search for suitable observables. One promising choice for the investigation for example of neutron-rich nuclei is to look after their excitation modes and trying to find correlations e.g. of transition probabilities and excitation energies with the size and shape of the neutron skin. The appearance of low-energy electric
dipole strength is a genuine feature of
neutron-rich nuclei, seen recently in high-precision photon scattering
experiments already in stable nuclei with small
\cite{Ca-pdr,Ba-pdr} and moderate \cite{Pb-pdr} neutron excess.
These so-called {\it Pygmy Dipole Resonances (PDR)} are located
close to the neutron threshold, forming in presently accessible
medium- and heavy-mass nuclei a clustering of states in the
region of E$_x\sim5.5\div$8~MeV. Although carrying only a small
fraction of the full dipole strength these states are of
particular interest because they are reflecting the motion of the
neutron skin against the core of normal nuclear matter. As
discussed in \cite{Pb-pdr}, their nature is completely different
from the representation where proton and neutron
fluids as a whole move against each other, giving rise to the
prominent isovector Giant Dipole Resonance (GDR).

The PDR mode has to be distinguished from the other known
low-energy isoscalar dipole excitation, namely the two-phonon
$1^-$ states resulting from the anharmonic interactions of the
lowest $2^+$ and $3^-$ states in a nucleus. A comprehensive
collection of data on the latter and overview of the widely used
experimental methods is given in \cite{Pitz}. The anharmonicities
are reflecting the intrinsic fermionic structure of the nuclear
phonons thus deviating from ideal bosons. A model taking into
account that nuclear phonons are formed by exciting at least two
quasiparticles (corresponding to a one particle one hole state (1p1h)) which
interact by a residual two-body NN-interaction is the
QPM \cite{Sol,Sol2} worked out and applied systematically \cite{Vdo,Vor,Gal}, e.g. also in the recent PDR
investigations in the stable $^{116,124}$Sn\cite{Pon1} and $^{208}$Pb \cite{Pb-pdr} nuclei. Other
applications of the QPM to low-energy dipole strength
\cite{Pon2,Gri1,Pon3} have led to very good description of data
thus giving confidence on the reliability of the model for such
investigations.

Most likely, the QPM approach is at present the only available
method for a unified description of low-energy single and multiple
phonon states. The use of schematic interactions of
separable form and with empirical coupling constants is a detail
which in principle could be improved on. Recently, finite rank 
approximation for Skyrme interaction was incorporated in the 
formalism of QPM \cite{Sev}. Work in this direction is
in progress but since the investigations in this paper are mainly
directed to an exploratory study of the evolution of the PDR modes
with neutron excess we use the standard form of the QPM.

Here, the mean-field part, however, is treated microscopically by
incorporating HFB results on single particle energies and potentials as
input for the QPM calculations. 

The interest on the region around the N=82 shell closure has
arisen recently \cite{Cor,Ter,Paar} and will further increase with the
new experimental opportunities on REX-ISOLDE, MAFF and also the
coming-up GSI project. In this manuscript we have presented calculations for the
neutron-rich Sn isotopes with neutron number N=$70\div82$ as an
exploratory investigation in experimentally less or even unknown
regions. 

\section{Description of the method}\label{Theory }
For the extrapolation of the QRPA and QPM calculations into unknown mass regions a reliable description of the ground state properties is necessary. 
Because of the numerical constraints set by the QPM a
semi-microscopic approach is chosen. The model Hamiltonian is
written as:
\begin{equation}
{H=H_{MF}+H_M^{ph}+H_{SM}^{ph}+H_M^{pp}} \quad ,
\label{hh}
\end{equation}
where $H_{MF}=H_{sp}+H_{pair}$
represents the mean-field part describing the motion of
independent quasiparticles in a static potential and interacting
in the particle-particle (p-p) channel through the monopole
pairing interaction  $H_{pair}$. $H_{MF}$ is to be identified with
the HFB Hamiltonian discussed in \cite{Hofmann,Len}. For practical purposes,
however, we use here a re-parameterization of the HFB mean-field in
terms of Wood-Saxon potential- $U_{WS}$ fitted to the
HFB mean-field such that the single-particle energies are reproduced. Also, the pairing part is simplified by using the
conventional constant matrix element approach which close to the
Fermi level is a reliable approximation.

$H_M^{ph}$ and $H_{SM}^{ph}$ are residual interactions taken as a
sum of isoscalar and isovector separable multipole and spin-multipole interactions in the particle-hole channel. 
$H_M^{pp}$ is the sum of the multipole
pairing interactions in the particle-particle (p-p) channel.

In practice, for a given nucleus of mass $A$ the depth of the
central and spin-orbit potentials, radius and diffusivity
parameters of $U_{WS}$ are adjusted separately for protons and
neutrons to the corresponding single particle separation energies,
the total binding energy \cite{Audi95}, the charge radii and
(relative) differences of proton and neutron root-mean-square
(RMS) radii,
\begin{equation}
\delta r=\sqrt{<r^2>_n}-\sqrt{<r^2>_p}
\end{equation}
indicating the thickness of the neutron skin. The results from this
procedure are compared to
experimental values from the compilation of Audi and Wapstra
\cite{Audi95} and the recent investigations of neutron skins by
Krasznahorkay et al. by charge exchange reactions \cite{Sn-skin},
respectively. For both observables the agreement of theory and
data is very satisfactory as it is seen from Fig.1.

\begin{figure}[htb]
\includegraphics{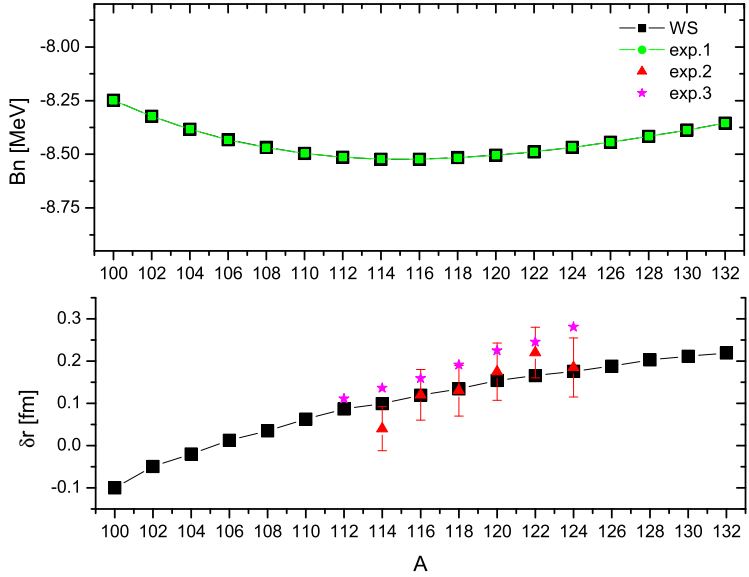}
\caption{\label{fig:fig1} Ground state properties of the Sn
isotopes. In the upper panel the total nuclear binding energies
per particle are compared to the values from the Audi-Wapstra
compilation \protect\cite{Audi95}. In the lower panel the
differences of proton and neutron rms radii are compared to the
experimental values obtained from charge exchange reactions in
ref. \protect\cite{Sn-skin}.}
\label{fig:fig1}
\end{figure}
The model Hamiltonian is diagonalized assuming a spherical $0^+$
ground state which leads to an orthonormal set of wave functions
with good total angular momentum JM. For even-even nuclei these
wave functions are a mixture of one-, two- and three-phonon
components in the following way \cite{Gri1}:
\begin{equation}
\Psi_{\nu} (JM) =
 \left\{ \sum_i R_i(J\nu) Q^{+}_{JMi}
\right.
\label{wf}
+ \sum_{\scriptstyle \lambda_1 i_1 \atop \scriptstyle \lambda_2 i_2}
P_{\lambda_2 i_2}^{\lambda_1 i_1}(J \nu)
\left[ Q^{+}_{\lambda_1 \mu_1 i_1} \times Q^{+}_{\lambda_2 \mu_2 i_2}
\right]_{JM}
\end{equation}
\[
\left.
{+ \sum_{_{ \lambda_1 i_1 \lambda_2 i_2 \atop
 \lambda_3 i_3 I}}}
{T_{\lambda_3 i_3}^{\lambda_1 i_1 \lambda_2 i_2I}(J\nu )
\left[ \left[ Q^{+}_{\lambda_1 \mu_1 i_1} \otimes Q^{+}_{\lambda_2 \mu_2
i_2} \right]_{IK}
\otimes Q^{+}_{\lambda_3 \mu_3 i_3}\right]}_{JM}\right\}\Psi_0
\]
where R,P and T are unknown amplitudes, and $\nu$ labels the
number of the  excited states.

The model basis includes the natural parity
1$^{-}$, 2$^{+}$, 3$^{-}$, 4$^{+}$, 5$^{-}$ phonons only. For the
description of the structure of each of the excited states we use
a wave function from Eq.~(\ref{wf}). For the case of 1$^{-}$ states
one-phonon configurations up to E$_x$=20~MeV are included.
This allows us to take into account microscopically the influence of core polarization on the low-lying 1$^{-}$ states without any phenomenologically induced effective charges. The two- and three-phonon configurations are
truncated up to 4.5 MeV for the calculation of the
quadrupole-octupole  1$^{-}$ state for the proper comparison with
the available NRF data \cite{Pon2}. For the QPM calculations
between 4.5$\div$8~MeV the two- and three-phonon basis is limited
to states up to E$_x$=8.5~MeV and 8~MeV, respectively. 

\section{Results}
The new method we have applied for the
determination of the ground state properties allows to investigate
the evolution of the dipole strength function with the neutron excess. 

Of our particular interest are the one-phonon 1$^{-}$ states below the neutron threshold with excitation energies up to 8~MeV in $^{120\div126}$Sn and up to
7.5~MeV in $^{128 \div132}$Sn respectively. 
In fact, the first three 1$^{-}$ QRPA states enter in
this energy region. They are rather well separated from the other
higher-lying one-phonon 1$^{-}$ states by an energy gap of more
than 1.3~MeV. The first QRPA state contains
mainly a neutron two-quasiparticle ($\approx 99.7 \%$) component while the
mixing between different two-quasiparticle neutron and proton configurations
(the contribution of the protons is about 0.2$\%$ only) becomes
more important for the second and the third QRPA states. These
states we identify with the PDR which in the case of our QRPA
calculations is located in the energy region 5.9$\div$7~MeV.
 The obtained neutron structure of the
$1^{-}$ QRPA states is in agreement with the explanation of the
PDR in other nuclei obtained before by DFT \cite{CA}, RQRPA \cite{Paar}
and QPM\cite{Pb-pdr} theory.

 This remarkable stability of the wave functions is similar to what
is found in theoretical studies of the GDR. In medium and heavy
nuclei the model independent Thomas-Reiche-Kuhn limit of the
dipole EWSR is found to be almost exhausted already by RPA or QRPA
calculations, indicating the dominance of 1p1h or two-quasiparticle structures
and agreeing with experiment. However, we emphasize that the PDR
states are of predominantly isoscalar (or eventually mixed
isospin) character as already pointed out in \cite{Pb-pdr}. Hence,
the PDR states cannot be considered as belonging simply to the
low-energy tail of the {\it isovector} GDR. Rather, this dipole
modes are of a genuine character which cannot be deduced by
extrapolations from the GDR region.

As a general result we find a correlation of
the total PDR strength and the number of the neutrons (presented on Fig.2a). 
The value of the total PDR strength increases in going to the heavier tin
isotopes, up to $^{132}$Sn. At the same time the centroid energies
show an opposite behavior (Fig.2b). 

We have extended the QRPA calculations in terms of QPM, which allows us to investigate the properties of
the lowest-lying  two-phonon quadrupole-octupole  1$^{-}$ states 

\begin{center}
\begin{figure}[htb]
\includegraphics{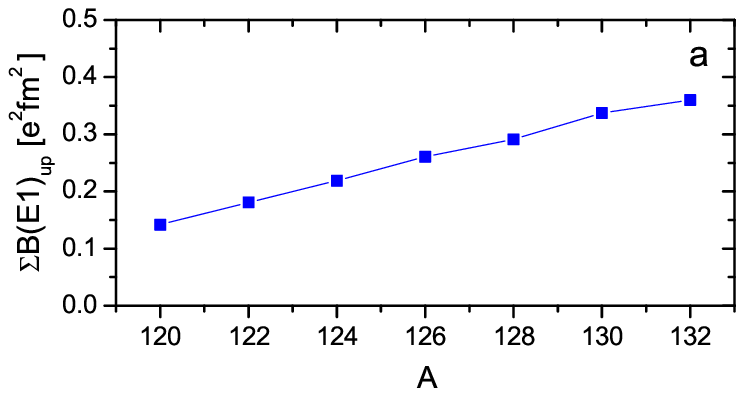}
\includegraphics{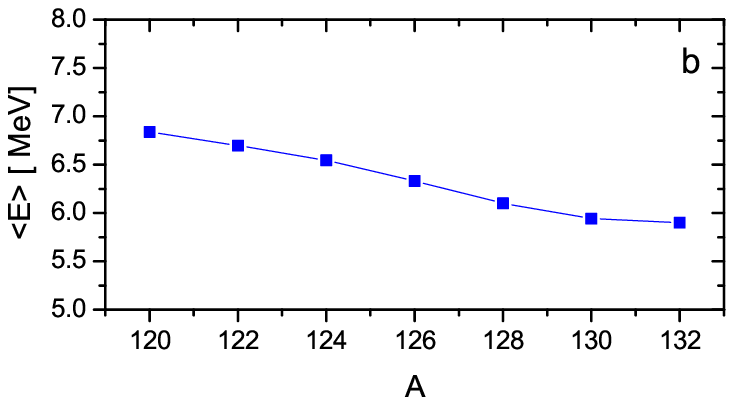}
\caption{\label{fig:fig1} QRPA calculations on the PDR a)total strength
and b)centroid energy as a function of the neutron access in $^{120\div132}$Sn isotopes.}
\label{fig:fig1}
\end{figure}
\end{center}
in $^{120\div130}$Sn isotopes as well. 

A proper description of the quadrupole
J$^{\pi}$=2$^{+}_{1}$ and octupole J$^{\pi}$=3$^{-}_{1}$ states is
important with respect to the two-phonon J$^{\pi}$=1$^{-}$ state.
The calculated energies and transition probabilities of the first 2$^{+}$,
3$^{-}$ states are presented in Table 1, where also 
the known experimental data are shown \cite{NDT}. A good agreement between the calculations and the experiment is obtained.
\begin{table}
\caption{\label{tab1}QPM results for the energies and the reduced B(E1), B(E2) and B(E3) transition probabilities 
of the first 1$^{-}$, 2$^{+}$ and 3$^{-}$ states in $^{120\div130}$Sn
isotopes. A comparison with the experimental data is presented \cite{Pon1}.}
\begin{tabular}{cccccccc}
\hline\hline
Nucl. &  & \multicolumn{2}{l}{Energy} & Trans. & &
\multicolumn{2}{l}{B(E1; I$_\nu ^\pi \rightarrow$ J$_{\nu'} ^{\pi'}$) [10$^{-3}    $ e$^2$fm$^2$]} \\ 
&  & [MeV] &  &  &  &\multicolumn{2}{l}{B(E2; I$_\nu ^\pi \rightarrow$ J$_{\nu'} ^{\pi'}$) [10$^4 $ e$^2$fm$^4$]} \\ 
&  &  &  &  &  &\multicolumn{2}{l}{B(E3; I$_\nu ^\pi \rightarrow$ J$_{\nu'} ^{\pi'}$) [10$^6 $ e$^2$fm$^6$]} \\ 
\hline
&J$_{\nu'} ^{\pi'}$ & Exp. & QPM & E$\lambda$ & I$_\nu ^\pi $ &  Exp. & QPM \\ \cline{2-8}
&  &  &  &  \\
$^{120}$Sn & 2$_1^{+}$ & 1.171 & 1.171 & E2 & 0$^{+}_1$& 0.200(3)& 0.193\\
&  &  &  & E1 & 3$_1^{-}$& 2.02(17) & 1.82\\
& 3$_1^{-}$ & 2.401 & 2.424 & E3 & 0$^{+}_1$& 0.115(15)
& 0.110\\
& 1$_1^{-}$ & 3.279 & 3.203 & E1 & 0$^{+}_1$&7.6(51)  & 7.6 \\
$^{122}$Sn & 2$_1^{+}$ & 1.141 & 1.137 & E2 & 0$^{+}_1$& 0.194(11) & 0.190\\
&  &  &  & E1 & 3$_1^{-}$&2.24(14)  & 2.06\\
& 3$_1^{-}$ & 2.493 & 2.486 & E3 & 0$^{+}_1$& 0.092(10)
& 0.099 \\
& 1$_1^{-}$ & 3.359 & 3.281 & E1 & 0$^{+}_1$&7.16(54)  &7.02  \\
$^{124}$Sn & 2$_1^{+}$ & 1.132 & 1.133 & E2 & 0$^{+}_1$& 0.166(4) & 0.174  \\
&  &  &  & E1 & 3$_1^{-}$& 2.02(16) & 1.98\\
& 3$_1^{-}$ & 2.614 & 2.645 & E3 & 0$^{+}_1$& 0.073(10)
& 0.087  \\
& 1$_1^{-}$ & 3.490 & 3.549 & E1 & 0$^{+}_1$& 6.08(66) & 6.27 \\
$^{126}$Sn & 2$_1^{+}$ & 1.141 & 1.151 & E2 & 0$^{+}_1$& -& 0.140  \\
&  &  &  & E1 & 3$_1^{-}$& - & 1.74\\
& 3$_1^{-}$ & 2.720 & 2.792 & E3 & 0$^{+}_1$& - & 0.079 \\
& 1$_1^{-}$ & - & 3.856 & E1 & 0$^{+}_1$& - & 5.8\\
$^{128}$Sn & 2$_1^{+}$ & 1.168 & 2.217 & E2 & 0$^{+}_1$& - & 0.097 \\
&  &  &  & E1 & 3$_1^{-}$& - & 1.07\\
& 3$_1^{-}$ & - & 2.849 & E3 & 0$^{+}_1$& - & 0.081 \\
& 1$_1^{-}$ & - & 4.115 & E1 & 0$^{+}_1$& - & 5.56\\
$^{130}$Sn & 2$_1^{+}$ & 1.221 & 1.204 & E2 & 0$^{+}_1$ & - & 0.066 \\
&  &  &  & E1 & 3$_1^{-}$& - & 1.11\\
& 3$_1^{-}$ & - & 2.861 & E3 & 0$^{+}_1$ & - & 0.098\\
& 1$_1^{-}$ & - & 4.094 & E1 & 0$^{+}_1$ & - & 5.53 \\
\hline\hline
\end{tabular}
\end{table}

The properties of the lowest lying 1$^{-}$ states have been
investigated in the frame of QPM. The energy varies from
E$_x$=3.203~MeV to E$_x$=4.115~MeV in $^{120\div130}$Sn (see Table 1). 
In these isotopes the structure of the 1$^{-}_{1}$ states is predominantly
by about 88$\div$93$\%$ of two-phonon quadrupole-octupole
character. Since with increasing neutron number the energy of the
latter becomes larger, certain other, higher-lying, two-phonon
configurations start to contribute as well. 
The three-phonon configurations are
most important for the investigated lower mass number Sn isotopes.
This effect is connected with the decrease of the
collectivity of the nuclear excitations when we come to the closed
N=82 shell.
The E1 transitions from the two-phonon 1$^{-}$ state to the ground
state and between 3$^{-}_1$ and 2$^{+}_1$ excited states are
an example of 'boson forbidden' transitions with one- or two-phonon exchange
\cite{Pon3}. The QPM calculations on the B(E1;
$0^{+}\rightarrow1^{-}_{1}$) and B(E1; 3$^{-}_1 \rightarrow 2^{+}_1$) transition probabilities are in a good agreement with the available experimental data. 
\begin{figure}[htb]
\includegraphics{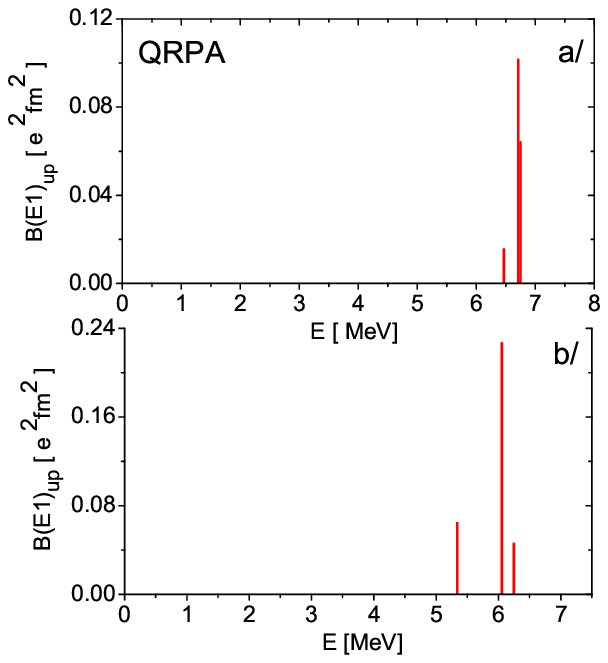}
\includegraphics{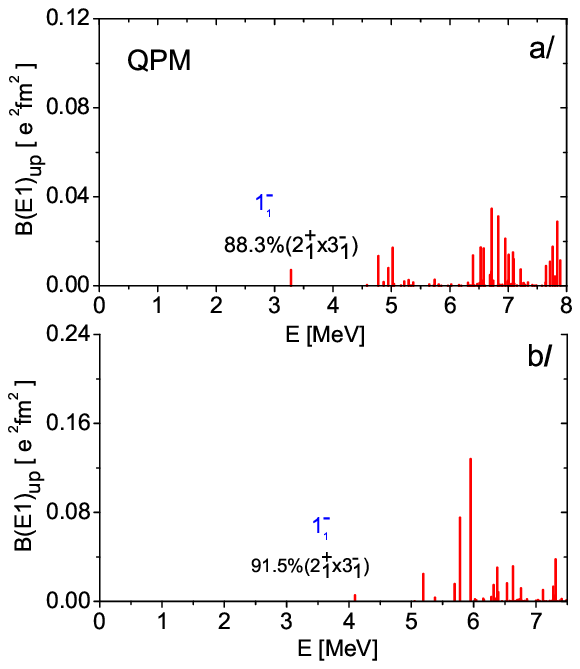}
\caption{\label{fig:wide}QRPA and QPM calculations of the E1
strength distribution in a/$^{122}$Sn and b/$^{130}$Sn below the
neutron threshold.}
\label{fig:fig1}
\end{figure}

A comparison between the QRPA and QPM dipole strength distribution
below the neutron threshold in $^{122}$Sn and $^{130}$Sn is
presented in Fig.3. The lowest QPM 1$^{-}$ state has no QRPA
counterpart because, as discussed above, it is of
quadrupole-octupole two-phonon structure. While in QRPA the dipole
strength in the investigated Sn isotopes is concentrated in three
dipole states the QPM phonon interactions lead to fragmentation -
in this case about 80 states - and dissipation seen as a
reduction of the total one-phonon content of the multi-phonon
eigenstates. \\
These quenching and fragmentation effects are directly observable
in the B(E1) transition probabilities to the ground state which
are determined by the one-phonon components of the excited states:
reductions in the one-phonon state amplitudes are reflected in a
corresponding suppression of the ground state transition matrix
elements. In $^{122}$Sn and $^{130}$Sn about 85$\%$ of the PDR
one-phonon strength is exhausted in the energy intervals up to
8~MeV and 7.5~MeV, respectively. Compared to the calculations in
\cite{Pon1} we find less fragmentation because the size of our
two- and three-phonon configuration spaces are somewhat smaller.
However, the total E1 strength in the investigated energy interval
is in very reasonable agreement with the available experimental
data in $^{124}$Sn. \\

\section{Summary and Conclusions}
A new method based on HFB and QPM theory was applied for the study
of the neutron-rich tin isotopes. From our calculations on
$^{120}$Sn $\div$ $^{132}$Sn we obtained low-energy dipole
strength in the energy region 5.9$\div$6.9 MeV concentrated in a
narrow energy interval such that a {\it pygmy dipole resonance}
(PDR) can be identified. The correlation of the PDR excitation
energy and transition strength with the neutron excess was
investigated. Its strength increases with the neutron number while
the centroid energy decreases. By comparison to the radii of HFB
ground state densities a correlation with the size of the neutron
skin could be identified.
An important step in understanding the dipole spectra is
to disentangle the PDR states from the low-energy two-phonon dipole
states. In our calculations this was achieved by using the QPM
approach with up to three-phonon configurations. 

For the experimentally unknown nuclei the excitation energy of the
two-phonon 1$^{-}$ state and B(E1) transition probability are
predicted. The method will be applied for other regions of nuclei.

The authors of the paper wish to acknowledge the discussions they
had with U. Kneissl and H.H. Pitz. This work is
supported by DFG, contract Le439/5.

\end{document}